# WAFER: A new method to retrieve sun-induced fluorescence based on spectral wavelet decompositions

Veronika Oehl [a,*], Alexander Damm [a,b]

[a] Department of Geography, University of Zurich, Winterthurerstrasse 190, 8057 Zurich, Switzerland
[b] Eawag, Swiss Federal Institute of Aquatic Science & Technology, Surface Waters – Research and Management, Ueberlandstrasse 133, 8600 Duebendorf, Switzerland

## A R T I C L E  I N F O

## A B S T R A C T

Sun-induced fluorescence (SIF) as a close remote sensing based proxy for photosynthesis is accepted as a useful measure to remotely monitor vegetation health and gross primary productivity. It is therefore important to develop methods that allow for its precise and reliable retrieval from radiance measurements with spectral resolutions that have been increasing over the past few years. Retrieval methods are catching up to the increasing complexity of the available datasets making use of their whole information extent (spectral, spatial and temporal) but the comparability of different SIF retrievals and consistency across scales is still important.

In this work we present the new retrieval method WAFER (WAvelet decomposition FluorEscence Retrieval) based on wavelet decompositions of the measured spectra of reflected radiance as well as a reference radiance not containing fluorescence. By comparing absolute absorption line depths by means of the corresponding wavelet coefficients, a relative reflectance is retrieved independently of the fluorescence, i.e. without introducing a coupling between reflectance and fluorescence. The fluorescence can then be derived as the remaining offset. This method can be applied to arbitrary chosen wavelength windows in the whole spectral range, such that all the spectral data available is exploited, including the separation into several frequency (i.e. width of absorption lines) levels and without the need of extensive training datasets.

At the same time, the assumptions about the reflectance shape are minimal and no spectral shape assumptions are imposed on the fluorescence, which not only avoids biases arising from wrong or differing fluorescence models across different spatial scales and retrieval methods but also allows for the exploration of this spectral shape for different measurement setups.

WAFER is tested on a synthetic dataset as well as several diurnal datasets acquired with a field spectrometer (FloX) over an agricultural site. We compare the WAFER method to two established retrieval methods, namely the improved Fraunhofer line discrimination (iFLD) method and spectral fitting method (SFM) and find a good agreement with the added possibility of exploring the true spectral shape of the offset signal and free choice of the retrieval window. On our synthetic dataset, WAFER seems to outperform the SFM and works best in a spectral window only containing solar Fraunhofer lines where we achieve a relative retrieval error of 10% on average. Applied to the real dataset, the method returns reasonable diurnal cycles for SIF and can, due to the decoupling of reflectance and fluorescence retrieval, reveal interesting trends at times when vegetation canopies may experience a midday depression that remain largely unobserved with current methods.

## 1. Introduction

Large scale observations of Earth, especially of vegetation, play a crucial role in gaining a better understanding of ecosystem processes and their adaptability to global environmental changes (Running et al., 1999; Schimel et al., 2019). In this context, carbon and water fluxes are of particular interest as they are highly sensitive to changing abiotic factors commonly associated with climate change such as precipitation or temperature (Reichstein et al., 2013). While carbon and water exchange between the ecosystem and the atmosphere can be quantified on a local spatial scale using eddy covariance towers (Baldocchi, 2020), remote sensing techniques allow for the retrieval of complementary information on abiotic and biotic factors to estimate and compare carbon uptake or evapotranspiration dynamics on larger scales. By exploring the interaction of solar radiation with the atmosphere, the surface and the ecosystems themselves, proxies to quantify the aforementioned processes can be obtained (Xiao et al., 2021).

Retrieving the desired flux dynamics occurring within and through vegetation from spectrally highly resolved intensities measured above the ground or from the top of the atmosphere requires a detailed understanding of the different contributing components. While various approaches based on evaluating the fractions of radiation being reflected by the surface have been developed and applied in the past to potentially quantify photosynthesis and related carbon exchanges, newer research focuses on the emitted sun-induced chlorophyll fluorescence (SIF, note that we will be using SIF and fluorescence interchangeably) signal. This emission of electromagnetic radiation can be observed in the red and near infrared and, being a byproduct of photosynthesis, has recently been found as a way to remotely access information about actual photosynthetic activity of vegetated ecosystems across different spatial scales (Mohammed et al., 2019; Ryu et al., 2019). Eventually, quantifying this signal allows for estimates of a number of related parameters and processes such as water (i.e. transpiration Damm et al., 2021; Pagán et al., 2019; Qiu et al., 2018) or carbon uptake (i.e. gross primary productivity (GPP) Damm et al., 2015; Guanter et al., 2014).

It has been demonstrated that SIF intensity and spectral shape at canopy level contain information about functional responses of plants (such as the tradeoff between photosynthesis, fluorescence and heat dissipation plants can regulate) to environmental conditions (Porcar-Castell et al., 2014) and are also sensitive to canopy structure (Fournier et al., 2012). The canopy architecture for example influences the far-red signal (Fournier et al., 2012), while a change in the plant condition causes changing contributions of photosystem I and II, leading to a difference in the ratio of the two peak heights (Joiner et al., 2016; Verrelst et al., 2016). This ratio and the respective emission intensities can also be sensitive to nitrogen uptake, which in turn influences the plants ability to fixate carbon (Corp et al., 2010).

Due to the small signal intensity (typically 1 % to 2 % of the total upwelling radiance in the NIR), retrieving SIF accurately with minimum prior assumptions and auxiliary data is one of the key challenges in the SIF remote sensing community.

Ignoring atmospheric absorption and scattering in between the surface and the sensor for a moment, the relationship between down- and upwelling radiance is quite simple: a fraction of sunlight gets reflected by the surface of the Earth and reaches the sensor at a certain distance above the ground. The SIF signal is independent of this process and simply adds onto the overall at-sensor radiance. This can be summarized as

$$s(\lambda) = R(\lambda)s_0(\lambda) + F(\lambda), \tag{1}$$

where $s$ is the upwelling at-sensor radiance, $s_0$ a reference spectrum (e.g. the downwelling top of canopy (ToC) radiance, more information on this can be found in Appendix B) and $R$ and $F$ denote reflectance and fluorescence respectively. Using spectroscopy, all of these quantities can be obtained at different wavelengths and are therefore equipped with a wavelength $\lambda$ argument.

Current retrieval methods rely on spectral shape assumptions for either reflectance or fluorescence or both, or are only using a fraction of the spectral channels available with state-of-the-art instruments, excluding a considerable amount of valuable information. So far, no schemes for the retrieval of SIF across the entire spectral range exist that are completely independent of spectral shape assumptions for the fluorescence, although recent developments relax such assumptions (Zhao et al., 2018). Aside from this, different SIF retrieval methods are used for measurement setups at different scales (i.e. in situ, airborne, satellite). This complicates a direct comparison of the retrieved fluorescence values, which is required for validation purposes (see Meroni et al., 2009 or Chang et al., 2020 for comprehensive overviews).

The FLD (Fraunhofer line depth or discrimination) methods represent the simplest way of disentangling the small fluorescence signal from the upwelling radiance. Eq. (1) is solved for the two unknowns $F$ and $R$ with two sets of measured values $s$ and $s_0$ making the simplifying assumption of $F$ as well as $R$ being constant across the wavelength range considered. Even though the method can in principle be applied to any two different points in the spectrum spectrally close enough, such that these simplifying assumptions hold, the FLD methods are mostly applied around the oxygen absorption lines where the absolute difference between two spectrally close radiances is large in order to reduce numerical errors and the influence of noise. This absolute difference can also be large enough in solar Fraunhofer lines, but in many cases the spectral resolution of the spectrometers used is not sufficient to measure the lowest point in the absorption line accurately. For some satellite acquired data, solar Fraunhofer lines can also be exploited (Joiner et al., 2011). The most widely used improvements to this simple approach are the 3FLD method (Maier et al., 2004) and the iFLD method (Alonso et al., 2008). Within the former, $F$ is still assumed to be constant but $R$ is modeled across the absorption band with a linear function, where the slope is approximated from the apparent reflectance $s/s_0$. Within the iFLD method, correction factors relating the true reflectance $R$, and eventually also $F$, outside and inside the absorption band are derived, again using the apparent reflectance to approximate $R$.

Approaching the problem similarly from Eq. (1) but using the whole spectral range instead of only a few distinct wavelengths is the spectral fitting method (SFM) (Cogliati et al., 2015, 2019). Here, $R$ is modeled as a piece-wise cubic spline with roughly 20 knots and $F$ as two Lorentzian functions with fixed peak positions multiplied with the reflectance (accounting for reabsorption) such that the two peak heights are the only free parameters for $F$. The multiplication with the reflectance induces a forced correlation between retrieved fluorescence and reflectance that does not necessarily reflect reality. The method takes $s$ and $s_0$ as inputs and then numerically minimizes the residual $(s(\lambda) - R(\lambda)s_0(\lambda) - F(\lambda))^2$ by varying the knot heights, i.e. $R$ values defining the spline, and the magnitude of the fluorescence peaks. This method is quite robust against noise but is limited to this predefined fluorescence shape and the residuals have to be carefully analyzed. The latter is important to make sure that the residual does not correlate with the distribution of the absorption lines which would indicate towards a mismatch of reflectance and fluorescence contributions to the overall signal as they act differently on the absorption line depths.

The singular value decomposition (SVD) method (Guanter et al., 2012) is based on decompositions of measured or simulated reference spectra not containing a SIF signal. The principal components describing most of the variance of these data are combined linearly to reconstruct a signal $s_0$ while the shapes of $R$ and $F$ are modeled as low order polynomials or are fixed and only modified by a constant. These retrievals are performed in selected wavelength windows where signature Fraunhofer lines are present. A drawback of the method is the dependence on the quality of the reference dataset needed to obtain the principal components. Past research successfully applied methods relying on reference datasets to satellite data (Guanter et al., 2012; Köhler et al., 2015). The application at local scales using airborne or drone data, however, can be challenging due to missing or heterogeneous reference datasets.

Within the Differential Optical Absorption Spectroscopy (DOAS) method the logarithm of the spectra is analyzed instead of the unaltered radiance values (Wolanin et al., 2015). This is a consequence of the method being developed to retrieve the column density of trace gases in the atmosphere based on the absolute line depths. Here, the logarithm is taken in order to retrieve optical thickness with a linear fit directly through the Lambert–Beer law. As Eq. (1) is already linear in the desired quantity $F$, this linearization operation does not seem to be a necessity for fluorescence retrievals. Moreover, products become sums inside a logarithm such that the technical advantage of exploiting the different effects of multiplication and addition on the line depths and shapes of the spectra is gone. If the spectral shapes are not tightly and precisely constrained, fluorescence and reflectance can become degenerate additions to the spectrum, which poses a disadvantage of this method. This especially holds when the fluorescence contribution is developed into a Taylor expansion, turning it into a smooth additional contribution to the forward model just like the reflectance. As both the reflectance and fluorescence are modeled as low order polynomials or with a fixed shape that could easily be described by a low order polynomial as well, assigning the two contributions to the respective processes becomes difficult.

As outlined in the last paragraphs, each of these aforementioned methods has successfully been applied to existing datasets and offers slightly different ways of retrieving SIF. The small signal in combination with many factors disturbing the retrieval requires and justifies the use of parametric modeling assumptions whenever possible. However, imposing models that do not describe the data well enough induces biases on the retrieved parameters that depend on the nature of the model. As different methods apply different models for the fluorescence in particular, the retrieved fluorescence values will differ depending on the chosen shape and different systematics will be introduced. A consistent way of retrieving fluorescence from different measurement setups is desirable in order to be able to directly compare SIF retrievals, e.g. in the context of validation activities (Buman et al., 2022). Apart from this, as datasets become more comprehensive, new ways to exploit all the information available need to be explored. Therefore, complementing existing retrieval methods with a new one that (i) makes use of all the spectral information available, (ii) does not make too many assumptions, (iii) is able to bridge the aforementioned gaps, and that (iv) is at best also scalable to different measurement setups would be an important step forward to advance the reliability of SIF retrievals.

The SIF retrieval method WAFER that will be introduced in this work is based on wavelet decompositions (WD) of the upwelling at-sensor radiance as well as a reference that can either be the ToC downwelling radiance, if known, a solar spectrum or a different upwelling at-sensor radiance from a non vegetated pixel (see Appendix B). Wavelet decompositions have been used for the analysis of spectral data in astrophysics (Meiksin, 2000; Starck et al., 1997) or to assess spectral shapes of ground reflectances (Cheng et al., 2014) but we are not aware of any prior work using them to estimate and compare absolute absorption line depths in atmospheric radiative transfer.

The paper is structured as follows: We will describe the WAFER method and model assumptions for our retrievals in Section 2, where we will also introduce the simulated and measured datasets on which it will be tested. The obtained results are presented in Section 3 and discussed and compared to the performance of existing methods in Section 4. In Appendix A, some more information about the radiative transfer assumptions is provided.

## 2. Methods and data

### 2.1. Spectral shape assumptions

Retrieving SIF from upwelling radiance spectra requires some assumptions or constraints on the spectral shapes of key components that cause the changes to the downwelling radiance. With WAFER, the number and restrictiveness of these assumptions is greatly reduced as compared to other methods while still obtaining comparable results. In this section we briefly argue why we made these choices and how they affect our results before describing the method.

#### 2.1.1. Reflectance

In the limited wavelength range where the SIF emission is present, the reflectance of vegetated surfaces can be described by a smooth function of wavelength (Mohammed et al., 2019; Woolley, 1971). In an optimization process, the parametrization should on one hand be flexible enough to represent the true and a priori unknown shape but on the other hand also be simple and robust to avoid overfitting and limit the required computational power. This is especially important when not tightly constraining the fluorescence shape at the same time. The entire ToC reflectance shape needs to be flexibly adjustable as it can vary significantly with species (Woolley, 1971), fractional cover by leaves in a given footprint and plant status (see for example Jacquemoud and Ustin, 2019). It can accurately be described by a piece-wise cubic spline (e.g. Cogliati et al., 2015). Since this requires a spline with roughly 20 knots for the wavelength range considered for SIF retrievals meaning a high number of 24 degrees of freedom and therefore parameters to be optimized, we have limited our retrievals to spectral windows where a second order polynomial with only 3 degrees of freedom suffices to represent the reflectance. The retrieval windows employed are listed in Table 1.

#### 2.1.2. Fluorescence

The Kennard-Stepanov relation (Kennard, 1918; Stepanov, 1957) establishes a very general link between the ground-state absorption spectrum and emission spectrum of an excited atom, molecule or pigment. This has been shown to also be applicable to solutions of photosystem I (Croce et al., 1996). On a molecular level, the spectral shape of the fluorescence mirrors the absorption spectrum of the absorbing molecule. In principle, this leads to temperature dependent but known emission shapes for chlorophyll molecules. In practice, the interplay of many chlorophyll molecules, leaf structure and also the canopy structure make the ToC emission shape very hard to predict. For SIF retrievals, it is argued that the assumption of a fixed spectral SIF shape induces little error for retrievals at fixed spectral points (Fournier et al., 2012; Guanter et al., 2013; Joiner et al., 2013). We suggest that this claim can only be underpinned when comparing to retrievals from spectra containing simulated SIF signals, which themselves are based on assumptions of the spectral shape and have their limitations when it comes to representing all possibly observable SIF spectra.

On a similar note, Magney et al. (2019) conducted a study measuring the shapes of fluorescence spectra from different species in the laboratory and analyzed them using an SVD. They conclude that most of the variance can be explained with a single spectral shape, if ambient parameters are controlled such that for example the temperature is kept constant. This only allows for conclusions about in-situ observable fluorescence shapes in a limited way.

Currently, there is no method in place to retrieve the full spectral fluorescence shape without prior shape assumptions apart from active methods in the laboratory, again with controlled ambient parameters. Therefore, we decided to not prescribe any shape assumptions on the fluorescence for WAFER in order to make it possible to measure not only the magnitude but also the rough shape of the signal.

As described in the last section, we will be imposing minor shape restrictions on the reflectance by using a second order polynomial. Consequently, the fluorescence shape retrieved as the remaining offset in fact will have a third order polynomial shape, if the downwelling radiance can at large scale (i.e. spectral scales larger than the absorption lines) be described by a first order polynomial and the upwelling radiance by a second order polynomial in the respective limited wavelength window. For the retrieval windows used with our approach, that is not too much of a limitation to the observable fluorescence shapes but should be kept in mind.

### 2.2. Retrieval technique

The WAFER[1] method is based on a wavelet decomposition (WD) (extensive information on this topic can be found in Mallat, 2009) of the measured spectral radiances and the separation of changes to a reference spectrum into multiplicative (e.g. reflectance) and additive (e.g. fluorescence) as summarized in Eq. (1) and further explained in Appendix A. A key property of the WD being exploited is the preservation of the spectral location while decomposing a given signal into different scales (spectral frequencies). This way, it can not only be analyzed which spectral frequencies contribute to what extent but also where in the spectrum. In this section, we will introduce the WD and highlight properties important for the SIF retrieval scheme developed on the basis of this operation. We will then outline the steps needed to disentangle the desired fluorescence signal from the mixed at-sensor signal using this approach. The retrieval scheme is illustrated in a flowchart (Fig. 1).

---

[1] https://github.com/voehl12/WAFER

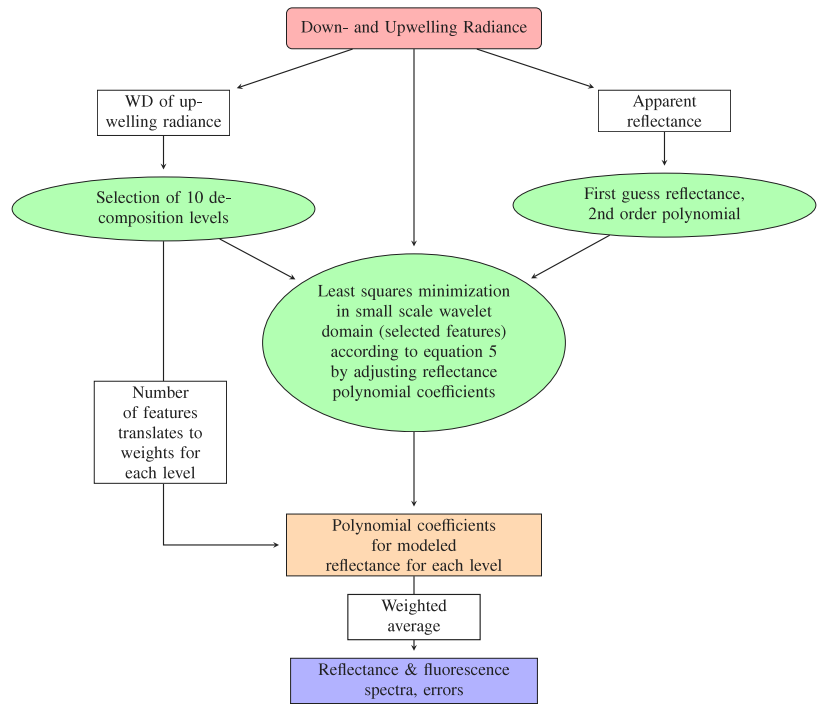

**Fig. 1.** Flowchart of the retrieval process. WD means wavelet decomposition.

### 2.2.1. The wavelet decomposition

The WD is the first step of the new SIF retrieval process and can be understood in analogy to a Fourier transformation (FT). In an FT, a signal $s(\lambda)$ is multiplied by a complex exponential function, that can be visualized by an infinitely extending sinusoidal function, with a given wavenumber and integrated over the real (wavelength) space to yield the signal strength at this isolated wavenumber or frequency in Fourier space. Similarly, in a WD, the signal is multiplied by a wavelet, a localized shape with adjustable width that determines the frequency, and integrated over the real space. The difference is that this shape is localized and can be pushed along the spectrum to yield the contribution of a given spectral frequency at a given spectral location, which makes it very useful for non-stationary signals, similarly to a windowed Fourier transform. Mathematically, the WD is a convolution of the signal with a wavelet with custom widths making it possible to detect and enhance different spectral features. This convolution can be expressed as

$$\hat{s}(\lambda, b) = \int d\lambda'\ s(\lambda')\ \psi\left(\frac{\lambda - \lambda'}{w_b}\right), \qquad (2)$$

where $\psi$ is the wavelet function and $w_b$ is the width parameter defining a decomposition level or scale. We will be using level and scale interchangeably, where low levels and small scales correspond to very narrow features and high levels and large scales represent spectrally smooth, slowly changing features. These wavelet functions are always meant to be normalized. The normalization factor depends on the width $w_b$ but is left out here for clarity since it is also unimportant for our application, as we will only be comparing wavelet coefficients of the same decomposition level. For brevity, we will write this convolution as

$$\hat{s}(\lambda, b) =: s * \psi_b. \qquad (3)$$

The convolved signal components are called wavelet coefficients and we will always refer to them as hatted quantities with a wavelength and width argument where needed. In order to make use of a WD, an appropriate wavelet function has to be selected from a range of possible shapes. For the continuous WD, only two mathematical characteristics are important: the functions have to be $L^2$-normalized and need to have a mean of zero.

As the continuous WD is a convolution of the wavelet function with the signal, one can expect the best representation of the input signal, when matching the wavelet shape to the expected signal shape. The WD will be used to detect absorption lines and quantify and compare their absolute depth, such that it is their shape that needs to be matched. As the approximate shape of the spectral response function of most spectrometers resembles a Gaussian, these observed absorption lines will also have a Gaussian shape. We therefore chose the widely used Mexican hat wavelet, which has a Gaussian-like shape being the second derivative of the Gaussian. It is possible to reconstruct the signal from the wavelet coefficients with this particular wavelet function with reasonably small residuals. Note though that a good reconstruction of the complete signal is not crucial for the method to work as the reconstruction is not needed in any of the retrieval steps. However, it is indicative of well chosen decomposition shapes and levels and therefore useful for selecting the function and decomposition scales.

For the task at hand, we employed the Python package PyWavelets (Lee et al., 2019) and, as this is not provided in the package, implemented a reconstruction for the continuous wavelet transformation following Torrence and Compo (1995) and Mallat (2009).

### 2.2.2. Level selection and weighting

Having selected the wavelet function, appropriate decomposition scales have to be chosen. This is shown as the first intermediate step on the left in Fig. 1. Upon a fine decomposition of the measured upwelling radiance into more than 2000 levels with widths between

**Table 1**

Retrieval windows and decomposition characteristics (shown for FloX-retrieval on 2021-04-23 and retrievals from synthetic data). All numbers are in nanometers and $b$ represents the wavelet scale width.

| Window | $b_{\min}$ | | $b_{\max}$ | |
|---|---|---|---|---|
| | FloX | SCOPE | FloX | SCOPE |
| 660–679 | 0.11 | | 0.24 | |
| 681–695 | 0.08 | | 1.11 | |
| 700–720 | 0.08 | | 0.31 | |
| 725–740 | 0.08 | | 0.29 | |
| 745–755 | 0.08 | 0.10 | 1.08 | 0.63 |
| 754–773 | 0.14 | 0.25 | 0.92 | 0.46 |
| 770–800 | 0.14 | | 0.43 | |

0.08 nm and 9 nm, the relevant scales are selected automatically by counting detected features. Features are defined as wavelet coefficients exceeding a level dependent threshold that depends on the median coefficient strength, which in turn serves as a proxy for the noise level on that scale. If the coefficient exceeds this median value times a factor (following soft thresholding outlined in Mallat, 2009) and is negative (i.e. represents an absorption line), it is counted as a feature. This way the range of levels used for the optimization can be narrowed down by only picking 10 levels with the highest numbers of features in a given wavelength range. For our FloX-data as well as the two retrieval windows for the synthetic data, we list retrieval windows and the corresponding selected wavelet scale widths based on this procedure in Table 1.

After the fit of the reflectance (see Section 2.2.3) has been performed for all selected levels, the same technique of counting features is applied to determine weights: the more features detected on a given level in a given wavelength range, the larger the weight when averaging the resulting reflectance polynomes from the different levels. This ensures that levels capturing more features are deemed more important and, as the fit is only performed with respect to these features, that a fit to more points is considered more reliable.

### 2.2.3. Separation of scales and optimization process

WAFER makes use of the separability of the spectral shapes of SIF (slowly changing over the spectral range, low spectral frequency) and the overall signal of reflected sunlight rapidly changing due to numerous absorption lines. As the fluorescence appears as an offset in the measured upwelling radiance (see Eq. (1)) and integration and therefore convolution is a linear operation, this contribution will vanish if the decomposition scales are chosen small enough. Everything affecting the downwelling solar radiation mutliplicatively (including but not limited to the reflectance) will influence the absolute line depths, which are captured and quantified by the small scale wavelet coefficients without having to select particular lines or features. The wavelet representations of typical down- and upwelling radiances are shown in Fig. 2. The chosen wavelength window contains only solar Fraunhofer lines, which become apparent as red dips in the decomposition coefficients. Comparing the overall strength of the wavelet coefficients in the top left and right panel of Fig. 2 for the down- and upwelling radiance respectively or explicitly comparing the two decompositions on level 5 (lower panel of Fig. 2), the impact of the reflectance on the absolute line depths becomes visually clear as a reduction in coefficient strengths.

We will retrieve the true reflectance or multiplicative alteration of the downwelling radiance by fitting a smooth function between the low level wavelet coefficients of the downwelling reference radiance and the measured upwelling radiance by minimizing the residual between the features in the wavelet space. The fluorescence will then be derived as the remaining offset. The third green item in the middle of Fig. 1 symbolizes this step.

The residual is calculated as the squared sum of the low level wavelet coefficients of the difference of the reference radiance times the modeled reflectance and the measured upwelling radiance:

$$\sum_{\lambda'} \left( (s - R s_0) * \psi_b \right)^2 \to 0. \tag{4}$$

Here, $\lambda'$ denote spectral positions counted as features according to the definition in Section 2.2.2 and $b$ denote the preselected levels. One important restriction imposed by setting the boundary conditions accordingly is that the term $s - R s_0$ cannot be zero by itself, otherwise the retrieved $R$ will simply be the apparent reflectance.

On each level $b$, Eq. (4) is minimized using the function minimize within Scipy's optimization package (Virtanen et al., 2020).

As in most optimization problems, it is advisable to give an initial guess and to constrain the result as well as possible, without making too many assumptions. In our case, the initial guess is given by the smoothed apparent reflectance, $R_{\text{app}} = s(\lambda)/s_0(\lambda)$, where the bump around the $O_2$A band is replaced by a smoothly continuing spline calculated as a second step in the scheme (Fig. 1, green item on the right). Boundary conditions are set such that the apparent reflectance minus a safety margin is the maximum value and the minimum is the apparent reflectance minus 0.2, corresponding to a SIF radiance of 20% of the reference, which is much greater than what SIF typically contributes to the upwelling radiance.

For each level, the minimization function will return the optimal polynome coefficients for the reflectance (orange box in Fig. 1), which are then weighted and averaged to a single reflectance as outlined in Section 2.2.2. Finally, the reference signal $s_0$ is multiplied with the retrieved reflectance and subtracted from the upwelling radiance. The remaining offset, calculated as

$$F(\lambda) = s(\lambda) - R(\lambda) s_0(\lambda) \tag{5}$$

represents the fluorescence and also contains any other kind of additive offset that does not alter the absolute line depth such as instrumental offsets. However, this limitation also holds for the FLD methods and all other methods where not all possible offsets are explicitly modeled with their particular shapes. Reflectance $R$ and fluorescence $F$ are the final output arrays as shown in Fig. 1, blue box at the bottom.

### 2.2.4. Error calculation

As described in Section 2.2.2, the final reflectance is calculated as a weighted average from the single level reflectance results. Consequently, it is possible to assign a weighted standard deviation as retrieval error for the reflectance giving an estimate of how well the derivations from different widths of spectral features agree. Additionally, if uncertainties for the measured or synthetic up- and downwelling spectra are known or can be estimated as statistical errors from the aggregation process, these can also be taken into account as $\Delta s$ and $\Delta s_0$. Following Eq. (5) and assuming all measurements to be Gaussian distributed, the errors from the measurement can be propagated onto the final retrieval as

$$\Delta F = \sqrt{\Delta s^2 + (s_0 \Delta R)^2 + (R \Delta s_0)^2}. \tag{6}$$

This error is also returned by the optimization function as shown in Fig. 1, blue box at the bottom.

### 2.3. Data

#### 2.3.1. Observational data

We test WAFER on a diurnal set of spectra on several cloudless as well as overcast days. These spectra have been acquired with the FloX-system (JB-Hyperspectral Devices, Düsseldorf, Germany) over an agricultural site in Oensingen, SO, Switzerland (47.2864° N, 7.7338° E). The site is situated on the Central Swiss Plateau, is exposed to temperate climate, and is characterized by crop rotation. A picture of the setup can be found in Appendix E. We use data from the year 2021 where the field was in a recovery period covered by grass that was regularly cut. This idealizes the conditions to test SIF retrievals as no significant shadowing effects are to be expected and the footprint of the spectrometer is mostly homogeneous. With this setup, upwelling as well as the corresponding downwelling radiances are available with a temporal sampling interval of 1 min to 2 min. According to the manufacturer, the full width at half maximum (FWHM) of the spectral response function amounts to 0.3 nm and the sampling rate is 0.17 nm. A signal to noise ratio of 1000 can be expected. To increase this signal to noise ratio, spectra have been averaged over a time period of 5 min resulting in a set of 154 spectra for a single day. Retrievals from single measurements are also possible but lead to slightly noisier diurnal cycles as one would expect.

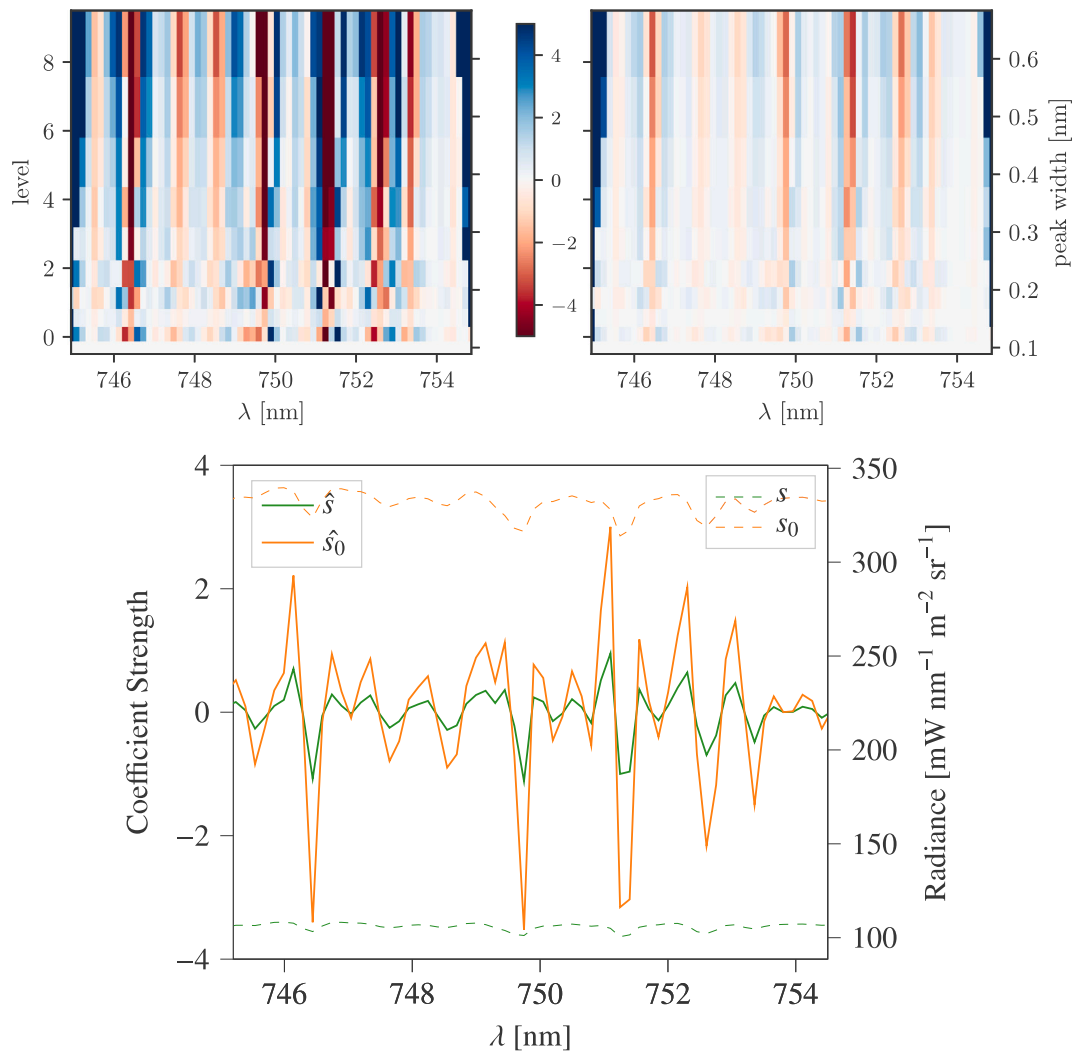

**Fig. 2.** Example of a wavelet decomposition of the downwelling radiance (upper left panel) and upwelling radiance (upper right panel). The colors encode the coefficient strength. For comparison both sets of coefficients for the downwelling radiance ($\hat{s}_0$, orange) and upwelling radiance ($\hat{s}$, green) on decomposition level 5 are also shown (lower panel). The corresponding radiance signals are shown as dashed lines. (For interpretation of the references to color in this figure legend, the reader is referred to the web version of this article.)

As systematical errors for this system are hard to determine (Buman et al., 2022) and are not of particular interest for this work, we only employ statistical errors in form of standard deviations calculated in the aggregation process.

The remaining noise is carried over to the wavelet decomposition, and, if the noise is uncorrelated across all wavelengths with a constant variance, will be present with the same power on all decomposition levels. Noise contributions are reduced within the retrieval by only selecting wavelet coefficients well above a noise threshold and averaging the retrieved reflectance over several decomposition levels. The assumption of certain noise characteristics is not important though, as the noise level is determined within the method for each decomposition level (see also Section 2.2.2).

### 2.3.2. Simulated data

To validate our retrieval method, we created a set of simulated ToC up- and downwelling radiance spectra. For a realistic representation of directional effects and the different contributions of direct and diffuse radiation, a directional parametrization for the reflectance has been employed for our numerical model. We made use of the option to calculate directional reflectance distributions with the SCOPE model, version 2.1 (van der Tol et al., 2009) and simulated a set of 56 bidirectional reflectance distribution functions (BRDF) and SIF spectra with varying chlorophyll contents (from 5 μg cm⁻² to 70 μg cm⁻² at intervals of 5 for the first interval and then 10 μg cm⁻²) and leaf area indices (from 1 m² m⁻² to 7 m² m⁻² at intervals of 1 m² m⁻²) leaving all other parameters at their default values and using a fixed solar zenith angle of 39.17° to match our observational data. Afterwards, the coefficients for a linear BRDF model have been obtained by inverting the linear equation numerically. The model chosen in our particular case is the semi-empirical Ross–Li BRDF discussed by Lucht et al. (2000) together with the hot spot extension by Maignan et al. (2004) as it is well tested and also implemented in the atmospheric radiative transfer code we employ, libRadtran (Mayer et al., 2017). The parametrization has been obtained by inverting the SCOPE BRDF sampled with 5° steps along both the viewing zenith angle and relative azimuth angle, with denser sampling around the hot spot. Once parametrized, i.e. after obtaining the kernels from Lucht et al. (2000), these reflectances can be extrapolated to any set of viewing angles, which libRadtran does internally. The wavelength dependence is also extrapolated internally, after the BRDF kernels are provided to libRadtran at the 1 nm sampling rate output by SCOPE.

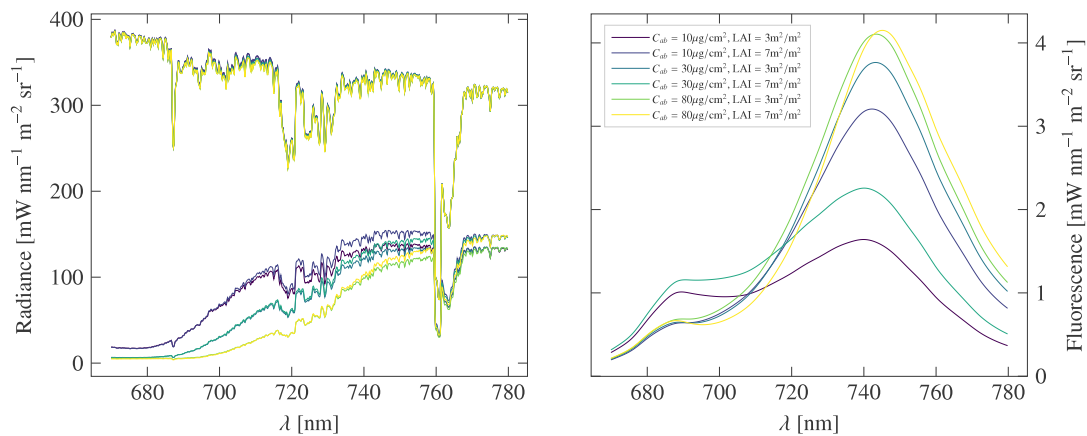

**Fig. 3.** Example up- and downwelling radiance spectra (left) as well as the modeled fluorescences (right).

For a leaf area index of $3 \, \text{m}^2 \text{m}^{-2}$ and a chlorophyll content of $50 \, \mu\text{g cm}^{-2}$, we show the BRDF for a selection of wavelengths in Appendix C (Fig. C.9) for the SCOPE simulation (producing reflectances for each set of zenith and azimuth angles at a wavelength sampling interval of 1 nm) as well as the parametrized version for comparison.

In Fig. C.10 in Appendix C we show the mean relative deviation of the parametrized bidirectional reflectance from the simulated one. This deviation does not exceed 3 % such that the parametrization seems to represent the simulated reflectances well enough.

The reflectances produced by SCOPE are not perfectly representable with this parametrization but it is the closest we can get to a realistic coupling of libRadtran and SCOPE without having to resort to extracting and separating multiplicative transmittances for the direct and diffuse parts of the radiation, as is done for example in Cogliati et al. (2019). This approach mostly overcomes the issue of the non-commutativity of spectral multiplication and convolution as the at-sensor radiance is calculated directly in the radiative transfer code, i.e. actually solving the full radiative transfer equation and including the coupling between atmosphere and bi-directional vegetation reflectance, and only tailored to the sensor specifications in the end.

Using libRadtran with a standard atmosphere and aerosols included, we created a set of down- and upwelling radiances at a sampling rate of 0.1 nm and a sensor height of 2 m above ground and again a solar zenith angle of 39.17°, replicating the set up as well as the location of our test site for 2021-05-30 at 10:00 UTC. The spectrally smooth SCOPE simulated SIF spectra (sampling rate of 1 nm) were interpolated and then resampled to match the libRadtran outputs. Then, they were multiplied with the transmittance between ground and sensor and added to the overall upwelling spectrum. Afterwards, the resulting highly resolved spectra were convolved with a Gaussian spectral response function with a FWHM of 0.3 nm and downsampled to the sampling rate of the FloX box (0.17 nm). A selection of the resulting spectra is shown in Fig. 3. Note that the BRDF also influences the observed downwelling radiance through the diffuse part of the radiation as it should be. This can be seen particularly well around 700 nm.

Additionally to this noise-free dataset, a second one with added noise has been used in our tests to assess the performance of WAFER on a more realistic but still synthetic dataset. Replicating mostly thermal noise of a typical CCD employed in a measurement setup as the one used for our observational data, we added Gaussian white noise with a signal to noise ratio of 1000 (see for example, the specifications and expected noise levels of the FloX-box, JB-Hyperspectral Devices, Düsseldorf, Germany).

For our particular dataset, 10 noisy spectra are averaged in each case. As most datasets are available as timeseries with a fairly high temporal resolution or spatially resolved pixels with similar characteristics in neighboring pixels, some kind of averaging by downsampling can realistically be applied on observational data as well.

## 2.4. Evaluation of WAFER

We evaluate WAFER by comparing retrieved fluorescences from simulated as well as observational data to the well established SFM (Cogliati et al., 2019) and iFLD method (Alonso et al., 2008) as described in Section 1. The following implementations and settings have been used:

For the SFM, we implemented a pipeline for our FloX-data using a least squares minimization with the Trust Region Reflective Algorithm (trf) of Scipy (Virtanen et al., 2020). For the modeling assumptions, we follow Cogliati et al. (2019) and model the fluorescence as two Lorentz distributions with fixed peak positions at 735 and 684 nm and corresponding scales of 25 and 10 nm respectively, multiplied by the reflectance. For the reflectance, we use a cubic spline with 4 spline knots for the same retrieval windows as used for our method for the synthetic data and additionally the whole spectral range with 24 knots for the observational data. As routinely done, the spectral regions of the apparent reflectance containing the oxygen absorption bands have been smoothed before inferring the reflectance. We would like to emphasize that using the whole spectral range as intended for the SFM resulted in worse results for the SFM on our synthetic data, which is why we left the results from the whole spectral range out for this direct comparison and discuss reasons for this behavior of the SFM in Section 4.1.

For the iFLD retrievals, we used an R implementation provided by JB-Hyperspectral Devices[2] specifically designed for the FloX-system. The implementation follows Alonso et al. (2008). In this application, the wavelength inside the absorption band is determined from the minimum downwelling and upwelling radiances between 755 nm and 765 nm or 682 nm and 692 nm for $O_2A$ and $O_2B$ respectively. The radiances outside the absorption band are chosen to be means evaluated in the range 4.12 nm to 3.12 nm towards smaller wavelengths starting from the selected wavelength inside the absorption band.

**Table 2**

Linear fit parameters and error measures comparing retrieved (using the spectral fitting method (SFM) and the new WAFER method) and modeled fluorescence values. These are shown for the simulated spectra with and without the addition of noise.

| Window [nm] | Method | Slope [−] | | Intercept [mW nm⁻¹ m⁻² sr⁻¹] | | $R^2$ [−] | | RMSE [mW nm⁻¹ m⁻² sr⁻¹] | |
|---|---|---|---|---|---|---|---|---|---|
| | | No noise | Noise | No noise | Noise | No noise | Noise | No noise | Noise |
| 745–755 | WAFER | 0.87 | 0.85 | 0.31 | 0.00 | 0.94 | 0.44 | 0.23 | 0.81 |
| | SFM | 1.14 | 1.17 | 0.27 | 0.23 | 0.91 | 0.68 | 0.75 | 1.02 |
| 754–773 | WAFER | 1.08 | 1.04 | −0.65 | −0.56 | 0.98 | 0.98 | 0.51 | 0.52 |
| | SFM | 1.02 | 1.02 | −0.55 | −0.55 | 0.99 | 0.99 | 0.52 | 0.52 |

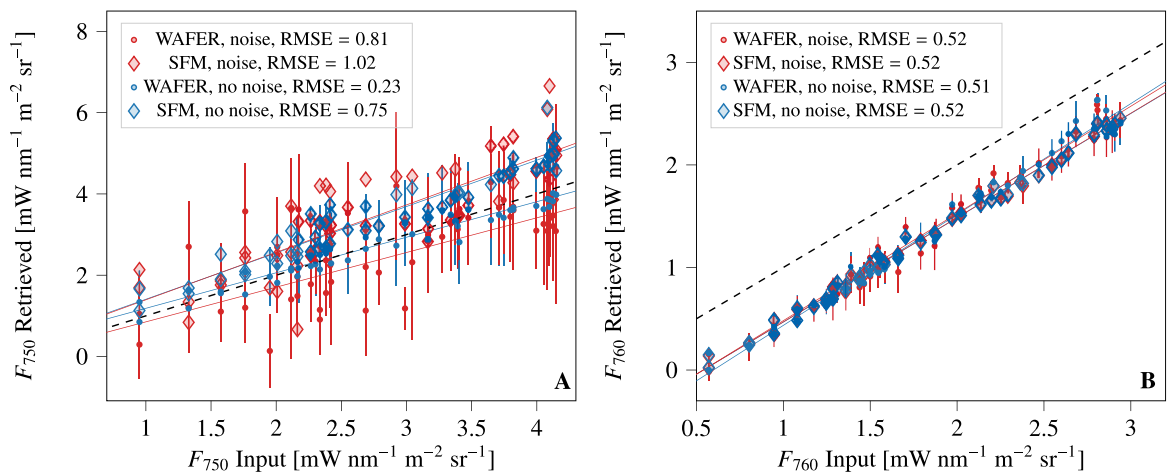

**Fig. 4.** Comparison of fluorescence at 750 nm ($F_{750}$) and 760 nm ($F_{760}$) between input from SCOPE and retrieved values for the new WAFER method (point markers and error bars) as well as the established spectral fitting method (SFM) (diamond markers). Results from two different retrieval windows are shown in the two panels: (A) using the window containing only solar Fraunhofer lines (745 nm to 755 nm), (B) using the window containing the atmospheric $O_2A$ absorption band (754 nm to 773 nm). Noisy (red) as well as noise free (blue) spectra have been used as inputs. The root mean square error (RMSE) is given in mW nm⁻¹ m⁻² sr⁻¹. The black dashed line represents the 1:1 line. (For interpretation of the references to color in this figure legend, the reader is referred to the web version of this article.)

## 3. Results

### 3.1. Evaluation of retrieval accuracy on synthetic data

The performance of WAFER as outlined in Section 2 and illustrated in Fig. 1 is evaluated using the synthetic spectra introduced in Section 2.3.2.

We compare the retrieved fluorescence values to the corresponding simulated SIF spectra, which is not possible with observational data where the true contribution of SIF to the upwelling radiance is not known. We focus on two retrieval windows: one containing only solar Fraunhofer lines (745 nm to 755 nm) assessed at 750 nm and a second one containing the atmospheric $O_2A$ absorption band (754 nm to 773 nm) assessed at 760 nm. For both windows, the retrieved SIF values are shown in Fig. 4 together with the errors calculated following Section 2.2.4 and the corresponding values retrieved with the SFM using the same retrieval windows. Additionally, the parameters of the linear fits and the root mean square error (RMSE) are listed in Table 2. In both retrieval windows, the RMSE suggests a better or similar performance of WAFER compared to the SFM for this particular synthetic dataset. For the solar Fraunhofer line window, the RMSE for WAFER amounts to 0.23 mW nm⁻¹ m⁻² sr⁻¹ without noise, corresponding to a relative root mean square error (RRMSE, relative to input values for each parameter set) of 10%, and 0.81 mW nm⁻¹ m⁻² sr⁻¹ with noise, corresponding to an RRMSE of 37%. For the SFM, the RMSE without noise amounts to 0.75 mW nm⁻¹ m⁻² sr⁻¹ and to 1.02 mW nm⁻¹ m⁻² sr⁻¹ with noise. The RRMSEs amount to 29% and 41%, respectively.

The dimensionless correlation measure $R^2$, with respect to the regression line, for WAFER is 0.94 and 0.44 for the noise-free and noisy spectra respectively and 0.91 and 0.68 with the SFM.

With RMSEs of 0.51 mW nm⁻¹ m⁻² sr⁻¹ and 0.52 mW nm⁻¹ m⁻² sr⁻¹ for WAFER and 0.52 mW nm⁻¹ m⁻² sr⁻¹ for both, the noise free and noisy input spectra for the SFM, the performance of WAFER is worse around the oxygen absorption window compared to the Fraunhofer line window, particularly for the noise free spectra, and also worse for the SFM, independent of the addition of noise. Here, all the RRMSEs amount to about 40%. The $R^2$ values are close to one for all retrieval methods for noisy as well as noise-free spectra.

We would like to emphasize that $R^2$ is only added because it is routinely done in the field not because we think it is a useful measure here. The regression line does not at all reflect on the performance of the method but is an indicator of linear correlation. $R^2$ is a measure of how much better a linear fit is than comparing to the mean of all measured values, which is not meaningful in this case where a correlation of input and measured values is something we would expect. $R^2$ is useful to determine the goodness of a linear fit or the strength of a linear correlation in order to find out how well the data is described by a linear relation. As we are not inferring (linear) fit parameters or trying to find a correlation here, $R^2$ is not relevant. If anything, it can tell us, how well constrained the offset, i.e. a systematic deviation of all retrievals from the input values, is or how far the data spreads around the regression line. The RMSE with respect to the model input is a better measure for the comparison of the performance of the methods and is therefore preferred. A good example for this problem is given in Fig. 4: in panel B, the linear regression is really good, $R^2$ is very close to one. But it is obvious, that the retrieved SIF values are strongly biased towards lower than the input values which is reflected in the RMSE. There seems to be a systematic deviation of the retrieved values from the input values that can with such a tight regression line be quantified with confidence to amount to roughly −0.6 mW nm⁻¹ m⁻² sr⁻¹ for all retrieval methods and independent of the addition of noise considering the intercept of these regression lines. Looking at the mean of the residuals would be a complementary way of evaluating the performance. This mean amounts to roughly $-0.5\,\mathrm{mW\,nm^{-1}\,m^{-2}\,sr^{-1}}$ for all methods as well as noise-free and noisy retrievals in panel B. In panel A, the mean of the residuals amounts to $-0.04\,\mathrm{mW\,nm^{-1}\,m^{-2}\,sr^{-1}}$ for the noise-free retrievals with WAFER and to $0.66\,\mathrm{mW\,nm^{-1}\,m^{-2}\,sr^{-1}}$ for the noise-free retrievals with the SFM. Adding noise, we see the same trend between the two methods with $-0.19\,\mathrm{mW\,nm^{-1}\,m^{-2}\,sr^{-1}}$ for WAFER and $0.70\,\mathrm{mW\,nm^{-1}\,m^{-2}\,sr^{-1}}$ for the SFM. Larger error bars especially for the noisy dataset indicate larger uncertainties for the retrieval in the Fraunhofer line window but a better agreement with the input values within this error.

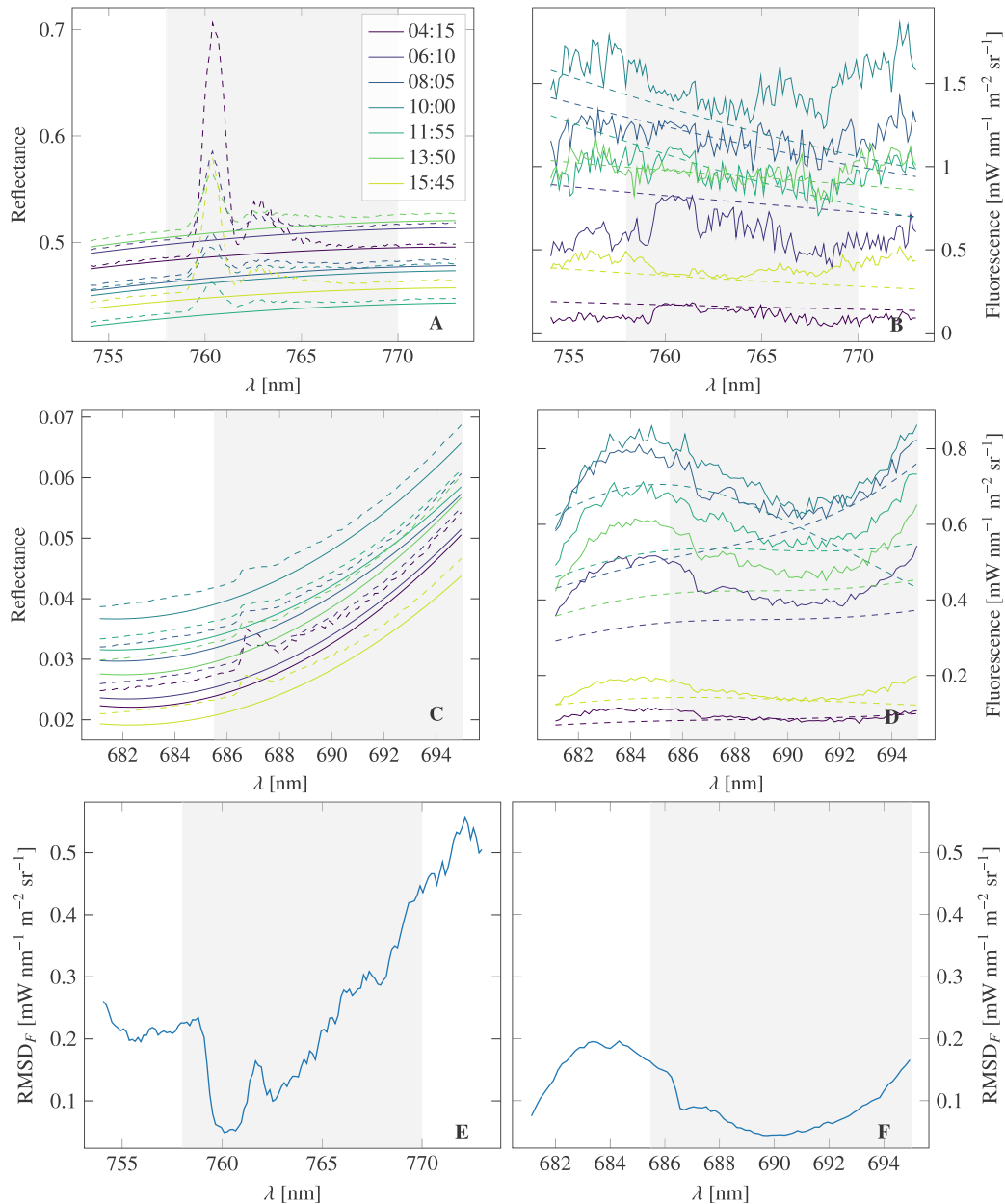

**Fig. 5.** Reflectances (A and C, solid lines) and fluorescences (B and D, solid lines) retrieved from measured radiance signals comparing reflectance retrievals using WAFER to the apparent reflectance and fluorescence retrievals to fluorescence retrieved with the established spectral fitting method (SFM) (dashed lines). Measurements were taken on 2021-04-23 and results are shown for the retrieval windows including either the O₂A band (A and B) or the O₂B band (C and D). The root mean square differences (RMSD) between SFM and WAFER retrieved fluorescences in these windows are shown in panels E and F, where the mean is taken over the entire day. In each panel, the oxygen absorption band is shaded in gray. (For interpretation of the references to color in this figure legend, the reader is referred to the web version of this article.)

### 3.2. Evaluation of retrieval accuracy on observational data

We applied the WAFER method to in-situ data acquired with a FloX-box as described in Section 2.3.1. For these data we looked at several retrieval windows: a larger window containing the O₂A absorption band, another one containing the O₂B absorption band and smaller ones connecting these two windows including one only containing solar Fraunhofer lines between 745 nm to 755 nm where no atmospheric absorption should be present. The windows have been chosen in such a way that the expected reflectance in each window is well represented by a second order polynomial and are listed in Table 1. In Fig. 5 we show retrieved reflectances and fluorescences for several times and two different retrieval windows on 2021-04-23. For reference, we also show the corresponding apparent reflectances as well as the SFM SIF retrievals as dashed lines. For these two retrieval windows, we also show the root mean square difference (RMSD) between our retrieval and the SFM for all wavelengths across all times of the day. This difference is noticably lower in the oxygen absorption bands and overall does not exceed $0.6\,\mathrm{mW\,nm^{-1}\,m^{-2}\,sr^{-1}}$ on this particular day. For low light conditions tested on 2021-05-11, both retrieval methods agree even better and the RMSD does not exceed 0.2 mW nm⁻¹ m⁻² sr⁻¹.

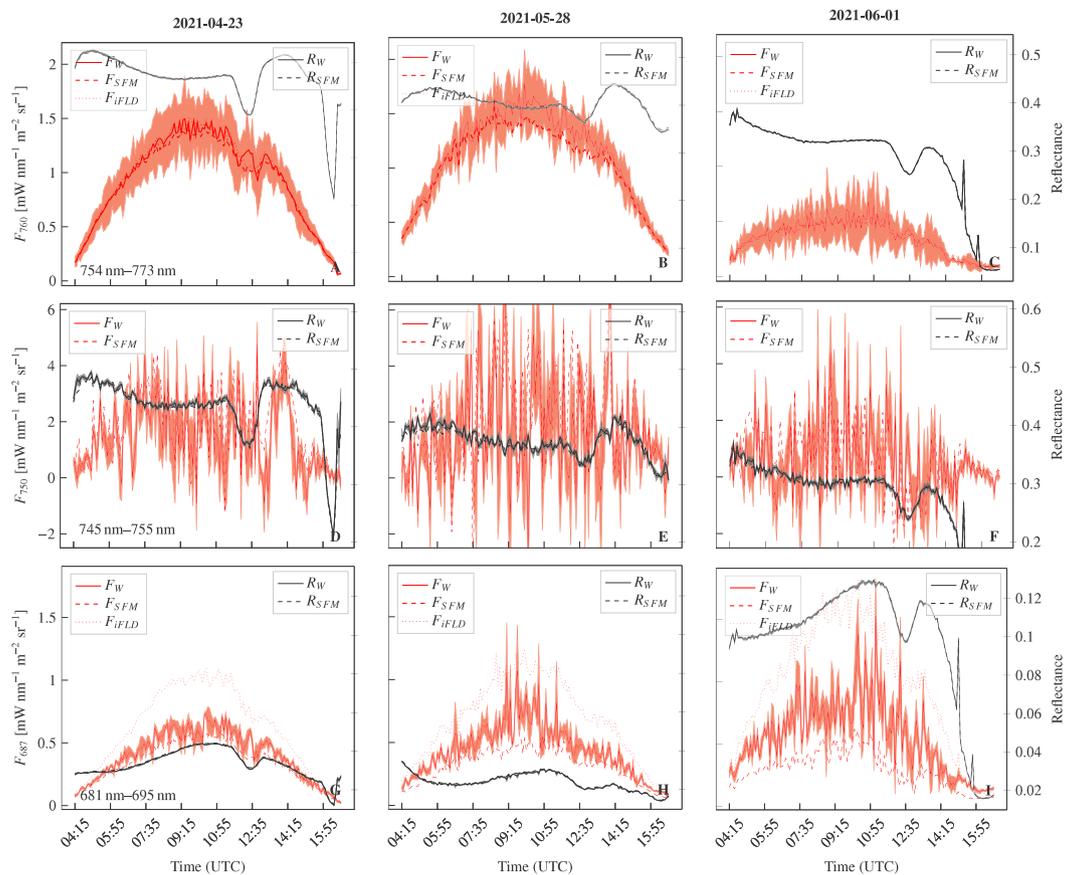

**Fig. 6.** Diurnal cycles of fluorescence and reflectance as derived with WAFER (solid lines) as well as the spectral fitting method (SFM) (dashed) and improved Fraunhofer line discrimination (iFLD) method where applicable (dotted). Different retrieval windows are shown in each row: 754 nm–773 nm with the fluorescence shown at 760 nm ($F_{760}$, A to C), 745 nm–755 nm with the fluorescence shown at 750 nm ($F_{750}$, D to F) and 681 nm–695 nm with the fluorescence shown at 687 nm ($F_{687}$, G to I). The data shown have been acquired with a FloX field spectrometer over a grass field on three different days (one column each): 2021-04-23 (A, D, G), 2021-05-28 (B, E, H) and 2021-06-01 (C, F, I).

On the overcast day (meaning low and completely diffuse light conditions), reabsorption of the fluorescence by oxygen becomes well apparent in the 754 nm to 773 nm retrieval window, especially for higher fluorescence values. This is shown in Appendix D, Fig. D.11.

### 3.2.1. Diurnal cycles

For the explicitly mentioned wavelength windows, we show diurnal cycles of the retrieved fluorescence at characteristic wavelengths together with the median reflectance in this window for three different cloudless days: 2021-04-23, 2021-05-28 and 2021-06-01 in Fig. 6. Here, fluorescence retrievals are only shown for a single wavelength in order to compare with the iFLD method, which relies on the oxygen absorption bands at 687 nm and 760 nm.

We also compare our results to the respective SIF values retrieved with the SFM, showing an overall agreement in the O₂A retrieval window, while the SFM tends to return lower SIF radiances in the lower wavelength ranges.

The retrieval from the 745 nm–755 nm window (i.e. only including narrow solar Fraunhofer lines) is much more noisy but the results show the potential feasibility of the new approach for a solar reference alone.

An interesting feature that becomes apparent in the O₂A retrieval window using WAFER is a decrease and increase in SIF just as the reflectance drops in the afternoon and again, when it increases roughly an hour later. The depression in the reflectance can also be seen clearly in the retrieval window including the red peak of the fluorescence signal situated around 684 nm (last row in Fig. 6) while the fluorescence only follows this trend when retrieved with the SFM or iFLD method.

### 3.2.2. Spectral shapes, full range retrieval

The spectral shapes in the left panel of Fig. 7 retrieved on 2021-04-23 are compared to the SFM shapes retrieved using the full wavelength range (note that the results for the synthetic dataset have been obtained using the same limited wavelength windows as for WAFER). This shows that using WAFER, SIF radiances can in principle be retrieved consistently across the entire wavelength range using the different retrieval windows independently. Employing the SFM in the same way, using only down- and upwelling radiances from a given wavelength window as inputs, only yields consistent results for some of the wavelength windows, particularly the windows containing the O₂ absorption lines, as the thin dotted lines in the right panel of Fig. 7 suggest. For comparison, the corresponding iFLD retrievals are also shown as diamonds.

The fluorescence around the O₂B band retrieved with WAFER is on average 0.15 mW nm⁻¹ m⁻² sr⁻¹ higher than the predefined SFM shapes allow (see also Fig. 5), albeit lower than the iFLD retrievals in this band. Note that the spectral shape of the fluorescence derived with WAFER depends on the model shape chosen for the reflectance and should not be mistaken as actual peaks of the fluorescence. It is not surprising that the shape segments look like third order polynomials, after the reflectance is modeled with a second order polynomial and the large scale wavelength dependence of the downwelling radiance is roughly linear in this small wavelength window.

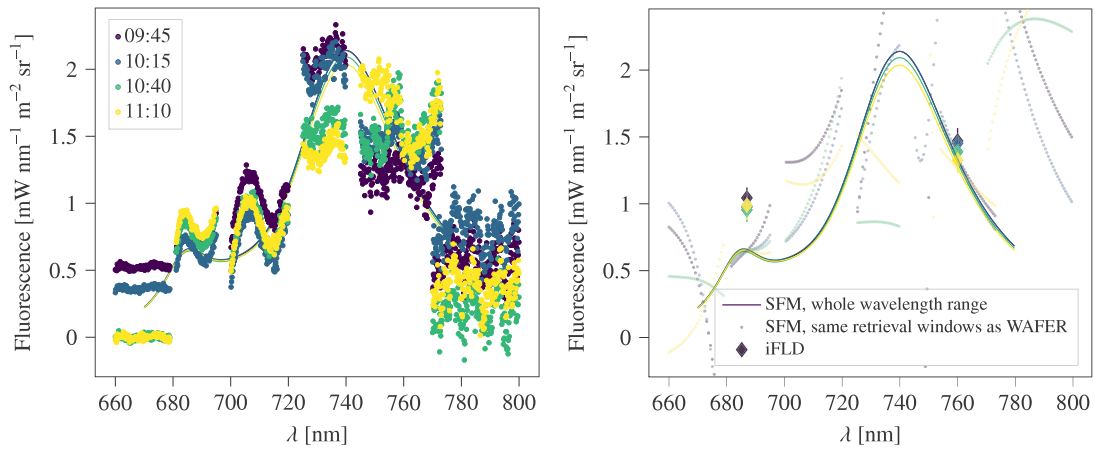

**Fig. 7.** Left panel: Full spectral range retrieved fluorescences for FloX field spectrometer measurements over grass at different times (UTC) on 2021-04-23 comparing the spectral fitting method (SFM) using the whole spectra (solid lines) and the new WAFER method using only limited retrieval windows (scatter). Right panel: SFM retrieved fluorescences using the full spectral range (solid lines) and using limited retrieval windows (thin dotted). Here, the improved Fraunhofer line discrimination (iFLD) method results are also shown as diamonds.

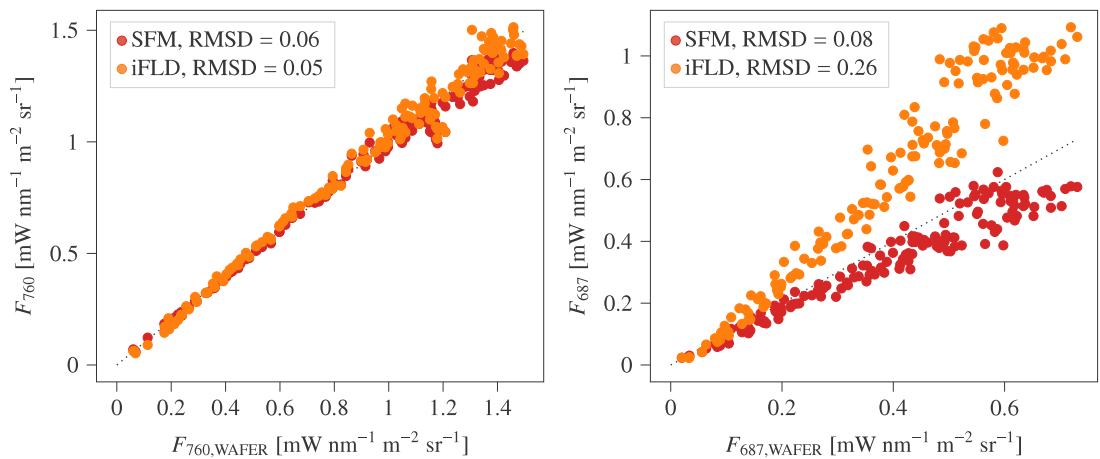

**Fig. 8.** Comparison of fluorescence retrieved using the improved Fraunhofer line discrimination (iFLD) method and spectral fitting method (SFM) against WAFER at characteristic wavelengths (left panel 760 nm ($F_{760}$) and right panel 687 nm ($F_{687}$)) for all measurements on 2021-04-23 over a grass field. Orange markers represent the iFLD method, red markers the SFM. The dashed lines denote the 1:1 line respectively. The root mean square differences (RMSD) are in $1\,\mathrm{mW\,nm^{-1}\,m^{-2}\,sr^{-1}}$. (For interpretation of the references to color in this figure legend, the reader is referred to the web version of this article.)

### 3.2.3. Direct comparison to other methods

In Fig. 8 retrievals with WAFER are directly plotted against the corresponding values retrieved with the SFM and the iFLD for all retrieval times on 2021-04-23. Here, we again use the same retrieval windows for the SFM as for WAFER and not the entire range for comparison. Note that in these particular retrieval windows shown here, the values retrieved with the SFM are the same irrespective of whether we use the full spectral range or only the small windows around the oxygen absorption bands. In the left panel the good agreement between SFM, iFLD and WAFER at 760 nm becomes apparent. The RMSDs between WAFER and the SFM and iFLD method amount to $0.06\,\mathrm{mW\,nm^{-1}\,m^{-2}\,sr^{-1}}$ and $0.05\,\mathrm{mW\,nm^{-1}\,m^{-2}\,sr^{-1}}$ respectively. For lower SIF values up to $1\,\mathrm{mW\,nm^{-1}\,m^{-2}\,sr^{-1}}$ the deviation does not exceed $0.1\,\mathrm{mW\,nm^{-1}\,m^{-2}\,sr^{-1}}$ for the SFM and $0.15\,\mathrm{mW\,nm^{-1}\,m^{-2}\,sr^{-1}}$ for the iFLD method. The relative difference to both other methods is less than 20 % across all retrieved values and 92 % of the retrievals on this day exhibit a difference of less than 10 % for both the SFM and iFLD methods. At least 90 % of all wavelet retrievals on this day have an absolute difference of less than $0.1\,\mathrm{mW\,nm^{-1}\,m^{-2}\,sr^{-1}}$ to both established methods.

In the $O_2B$ band, WAFER retrieves values roughly in between those that the SFM and iFLD methods return. This can be seen well in the right panel of Fig. 8. Here, the RMSD between the different methods is slightly higher ($0.08\,\mathrm{mW\,nm^{-1}\,m^{-2}\,sr^{-1}}$ and $0.26\,\mathrm{mW\,nm^{-1}\,m^{-2}\,sr^{-1}}$ for the SFM and iFLD method respectively). Again, the absolute agreement is better for lower fluorescence values but interestingly, the relative difference between WAFER and the SFM is roughly constant between 0 % and 30 % whereas this relative difference to the iFLD method increases towards higher fluorescence values. In both comparisons, the absolute difference does not exceed $0.1\,\mathrm{mW\,nm^{-1}\,m^{-2}\,sr^{-1}}$ for retrieval values less than $0.2\,\mathrm{mW\,nm^{-1}\,m^{-2}\,sr^{-1}}$ and 89 % of all absolute differences for the SFM and 26 % for the iFLD method are below $0.1\,\mathrm{mW\,nm^{-1}\,m^{-2}\,sr^{-1}}$. The corresponding plots for an overcast day can be found in Fig. D.12. In this case at 687 nm, both the iFLD method and the SFM agree better with the WAFER method with RMSDs of $0.03\,\mathrm{mW\,nm^{-1}\,m^{-2}\,sr^{-1}}$ in both cases and even better agreement at 760 nm.

## 4. Discussion

In the following we will discuss the performance as well as advantages and disadvantages of the WAFER method. Overall, we presented a method that is very flexible in its application and can in principle be used for in-situ as well as airborne and satellite acquired data. It does not rely on any spectral shape assumptions for the fluorescence and therefore allows for the exploration of different shapes that might arise in realistic measurement setups. We will discuss the performance of WAFER, where we only constrain the reflectance to be modeled by a second order polynomial, in comparison to well established retrieval methods that make assumptions or simplifications in Section 4.1. Apart from that, the retrieved fluorescence is not coupled to the reflectance, as the reflectance is extracted directly from the wavelet coefficients, as we will discuss in Section 4.2. WAFER decomposes each spectrum using wavelets and therefore exploits all the information that is available including the width of the absorption lines. With this comes naturally a higher computational cost, which we will discuss in Section 4.3 among other possible limitations.

### 4.1. Comparison of the new WAFER method to other methods

We applied the new wavelet based SIF retrieval method, WAFER, to synthetic as well as observational data and compared its performance to the established iFLD method and the SFM (see Chang et al. (2020) for an extensive comparison).

For the synthetic spectra, we specifically looked at two retrieval windows, one containing the $O_2A$ band, the other one only containing solar Fraunhofer lines. For the retrieval window including the $O_2A$ band, there is an obvious bias towards lower retrieved SIF values. Comparing the actual multiplicative function between down- and upwelling radiance without fluorescence that includes oxygen absorption effects in this wavelength range to the retrieved reflectance, the reason for this behavior can be explained: The absorption by oxygen continues as light travels from the canopy to the sensor (Sabater et al., 2018), such that the ratio of up- and downwelling radiance shows a dip around 760 nm. This dip cannot be modeled and fitted properly with a simple parabola. Therefore, the reflectance is always overestimated around this dip leading to fluorescence retrievals that are too low as can be seen in Fig. 4, right panel. Since this is a systematic error, the retrieval error bars for our method do not, but also cannot, account for this discrepancy. In this case, the SFM, which excludes the oxygen absorption band region to model the smooth reflectance and therefore is prone to the same error, and WAFER seem to agree better but in fact, both underestimate SIF. How strong this effect is with observational data is hard to determine, as the extent to how much oxygen absorption contributes on the way to the sensor is obscured by the addition of fluorescence and irregularities in the true reflectance. Nevertheless, a correlation of the spectral shape of our retrieved fluorescences with the shape of the oxygen absorption band is visible, which we will come back to later when discussing the observational results.

The retrievals from the solar Fraunhofer line window ranging from 745 nm–755 nm are more reliable. Due to the lack of atmospheric absorption (Frankenberg et al., 2012), a smooth reflectance without further transmittances is expected. Our SIF retrievals from the synthetic data for this wavelength range confirm this argumentation (i.e. by showing a lower RMSE and no clear bias) but it also becomes obvious that retrievals from this window are more sensitive to noise as the addition of noise immediately changes the wavelet estimated line depth of the very narrow solar Fraunhofer lines. This noise sensitivity can be accounted for to some extent by averaging the reflectance fit over several absorption lines and several decomposition levels.

Using WAFER, fluorescence retrievals from different wavelength ranges are completely independent such that the retrievals in the solar Fraunhofer line window are not influenced by the errors from the oxygen band windows. This offers new possibilities to compensate for atmospheric disturbances and complements latest developments to, for example, account for atmospheric re-absorption of SIF for in-situ measurement setups (Sabater et al., 2018; van der Tol et al., 2023).

Concluding the comparison to the SFM, it becomes apparent that the SFM is more robust against noise which was to be expected as the fluorescence shape is constrained, but WAFER shows an overall higher retrieval precision with our synthetic data. As retrieval methods have not been tested on similar synthetic datasets using BRDFs yet, more work is needed to tell whether our synthetic dataset does not represent real setups well enough or whether the SFM used with the parameters and settings as introduced in Cogliati et al. (2019) only works well on synthetic datasets modeled in the same way as in the corresponding publication.

Analyzing observational data measured with a field spectrometer as described in Section 2.3.1, we used several retrieval windows distributed along the entire spectral range of the instrument. Here, we were able to compare to the SFM as well as the iFLD method used on the same dataset.

Generally, all retrieval methods agree very well when SIF is retrieved in the $O_2A$ band. In the other retrieval windows, the SFM derived values are often slightly below our SIF retrievals or at the same level. This is especially true for the lower wavelength range including the $O_2B$ band (see Fig. 8). A possible explanation is that the SFM has the reabsorption of fluorescence implemented within the method as a multiplication with the reflectance function, which assumes low values in this wavelength range and therefore forces the red peak to be lower than it might actually be. Consequently, the reflectance would be over- and the fluorescence underestimated. A careful analysis of the SFM least squares residuals to disentangle multiplicative and additive changes to the spectrum of the incoming radiation could shed light on this and will be briefly touched upon later.

Not predefining the shape of the fluorescence signal allows for an analysis of the entire additional offset of the upwelling signal compared to the downwelling signal after having subtracted the multiplicative contribution. We can therefore see oxygen absorption contributions in our retrieved spectral shapes.

Notably the SIF retrievals in the far-red from an overcast day (see Fig. D.11, panel B) show an absorption pattern that looks like oxygen absorption in the optical path between sensor and canopy: it can either stem from the overestimation of the reflectance (or rather the fact of ignoring the transmittance) or be due to actual absorption of the fluorescence by oxygen, even though the latter phenomenon should have a negligible effect on the retrieved fluorescence within 2 m above the sensor as our transmittance radiative transfer models show. Most probably both, absorption of the reflected radiance and absorption of the emitted fluorescence, contribute. This shape similarity with the known oxygen absorption pattern is much stronger when the overall fluorescence is higher, as we would expect. It is interesting to see that the SFM and WAFER agree best in the oxygen absorption band (see Fig. 5, E and F), which hints towards the SFM mainly deriving the final shape of the SIF signal from this part of the spectrum. This can also be seen in Fig. 7, right panel, where the retrievals from the smaller wavelength windows only agree with the whole wavelength range retrievals in the oxygen absorption bands, i.e. the information coming from these windows seems to be determining for the whole fit, leaving the remaining parts of the spectral information essentially unused.

The fact that the SFM mainly relies on the oxygen absorption band can also be seen when looking at the residual of the SFM (not shown in this work): There is a consistently negative and clearly non-random residual at 687 nm, meaning that the forward modeled upwelling radiance is lower than the measured input speaking for an overall underestimation. Likewise, there seems to be a mainly negative residual in the Fraunhofer lines meaning that either the reflectance is over- or the fluorescence underestimated.

Meanwhile, Table 1 shows that the WAFER method does not exploit the large oxygen absorption features but instead relies on features much smaller in wavelength width.

Comparing WAFER to the iFLD method, a different observation consistent with the expected bias from looking at the retrieval equations (Alonso et al., 2008) has been made: in most of the cases tested, the iFLD returns higher values for the fluorescence as compared to the retrieval with other methods, which becomes particularly obvious in the $O_2B$ band (Fig. 8). This is due to the fact that the employed reflectance ratio is estimated from the apparent reflectance which can mathematically be shown to lead to an overestimation of the retrieved fluorescence. This effect is larger in the lower wavelength range because here the relative difference between the true reflectances within and left of the absorption band is larger and so is the error induced by using the apparent reflectance ratio instead. On the tested overcast day, the iFLD method in the $O_2B$ band returns more similar values compared to WAFER and does not exhibit as much of a clear trend. This might be due to an effect opposing the general inherent overestimation caused by the very low depth of the $O_2B$ band and therefore sensitivity to noise.

In the Fraunhofer line window, the general trend of the SIF diurnal cycle is visible with our retrievals but it is quite noisy. The reason for this is likely a combination of spectral resolution of the spectrometer and the signal-to-noise ratio that compromises resolving the depths of the Fraunhofer lines accurately enough. Also, for satellite based SIF retrievals operating in spectral regions where only Fraunhofer lines are present, an increased SIF retrieval noise has been reported (Guanter et al., 2021).

### 4.2. Phenomenological insights

With WAFER applied to in-situ data, a few interesting observations regarding fluorescence diurnal cycles and spectral shapes could already be made but they should not be taken as robust findings yet. Instead, this is to foreshadow what could be possible using this approach on more and different datasets.

The diurnal cycles, especially in the afternoon, seem to show a dependence on the retrieval method employed. With WAFER, it is possible to completely decouple the reflectance retrieval without having to approximate the shape of the fluorescence or the change of the reflectance along the retrieval window. As a result, we have seen fluctuations in the retrieved fluorescence in the afternoon before and after the reflectance drops. The fixed spectral shape of the fluorescence that is also multiplied with the reflectance in the SFM does not allow for these independent dynamics: when the reflectance decreases significantly in the respective wavelength range, SIF is lowered simultaneously. This is why the fluorescence follows the reflectance trend with the SFM but not with the newly developed WAFER method. With the iFLD, the fluorescence also follows the reflectance trend, but the coupling seems to be a bit weaker. This can also be explained to be coming from the method itself: The ratio of the reflectance outside and inside of the absorption band is estimated from the same ratio of the smoothed apparent reflectance (Alonso et al., 2008). This estimate does not necessarily change when the reflectance decreases but the fluorescence increases at the same time, whereas this ratio taken from the real reflectance might change. As the retrieved fluorescence values are very sensitive to this ratio, they might tend to follow the diurnal cycle of the apparent reflectance and are therefore not independent.

More generally, observed reflectance dynamics can be caused by plant physiology, but also by optical effects like reflectance anisotropies or shadowing (Kükenbrink et al., 2019). If the SIF retrieval method employed is coupled to the reflectance retrieval, the retrieved SIF dynamics might also not be related to photosynthetic regulation but to these optical effects. The different results using different retrieval methods shown here might be a starting point to develop SIF retrieval approaches less sensitive to optically caused reflectance dynamics (as for example discussed in Chang et al., 2021).

The sudden change in the retrieved range of fluorescence values between 2021-05-28 and 2021-06-01 (see Fig. 6) suggests that changes were made at the test site. In fact, the grass has been cut in between these two days, which immediately becomes obvious from the lower fluorescence and reflectance values around the $O_2A$ absorption band. Interestingly, there seems to be an increase in $O_2B$ fluorescence at the same time, which could be due to the fact that there is less reabsorption with shorter grass (Fournier et al., 2012). This effect is less obvious with the SFM but can be observed looking at the iFLD retrievals as well as the WAFER retrievals to a lesser extent. Meanwhile, the reflectance in this wavelength range increased by roughly 0.1 and decreased about the same amount in the $O_2A$ wavelength range.

The ability to observe different wavelength ranges independently essentially covering the entire range of a given spectrometer is a clear advantage of WAFER. In principle, the entire spectral range of the fluorescence can be explored at once without having to prescribe shape functions for the fluorescence by making use of all the spectral information available. With most established retrieval methods, the focus lies on the far-red range but it has been shown that the red range yields additional information (Joiner et al., 2016) as we were also able to demonstrate with the diurnal cycles.

### 4.3. Technical limitations

As mentioned in Section 2.2.3, WAFER only returns the additive offset (note that this is also true for all FLD methods), that contains SIF but can also be enhanced by instrumental effects. This is unlikely in the case of the FloX-retrievals because the fluorescence vanishes at the beginning and end of the day as it should, not showing any constant offset independent of the light conditions (Fig. 6). Also, the SFM with a prescribed fluorescence shape yields similar values in each wavelength range (Fig. 7).

Despite this limitation of the new WAFER method of not being able to distinguish different kinds of offsets, we found a good agreement with established retrieval methods for the in-situ data as well as reasonably looking diurnal cycles resembling the shape of a cosine with the maximum situated around solar noon. For airborne SIF retrievals, the problem of always retrieving SIF combined with instrumental offsets could easily be controlled by comparing to retrievals from non vegetated and therefore non-fluorescing targets (Damm et al., 2014; Siegmann et al., 2021).

Other inaccuracies in the retrieved SIF values could arise from fixing the reflectance to a parabolic shape, which might not model the true shape of the reflectance perfectly in each retrieval window. However, the effect should be negligible especially when handling the exact shape of the retrieved fluorescence with caution. The magnitude of the resulting fluorescence is not influenced by the particular shape of the reflectance when the retrieval window is small enough.

At this stage, the optimization process is relatively slow because the wavelet decomposition needs to be performed in each iteration, which can be quite intensive computationally (i.e. a factor of 50–60 slower than the SFM for FloX-data). We aimed at analyzing spectral data in new and potentially more holistic ways, which naturally comes with higher computational cost. Still, all retrievals presented in this work have been performed on a laptop and a single core within hours.

A significant speedup can be achieved by fitting the multiplicative function between the wavelet coefficients of the upwelling radiance and the reference directly, only decomposing once. This would mean pulling the reflectance out of the convolution integral in the second term of Eq. (4), which could be justified for very narrow low level wavelet functions such that the reflectance can be treated as a constant along the wavelength. Of course, this is a simplification that would lead to a loss in accuracy but might still be an option for initial estimates of observations that have high data volumes. Further research is needed to quantify the effect of this simplification.

## 5. Conclusion

The performance assessment of the newly developed wavelet based SIF retrieval method WAFER presented in this work suggests that it can be taken as an alternative or supplement to established retrieval methods. Combining basic principles from different retrieval methods, it makes use of all the spectral information available in a given wavelength window. The absolute line depth of different absorption lines is quantified like in all of the FLD methods, while the spectrum is also decomposed as in the SVD method but not to reduce the dimensionality but rather to filter for features and examine alterations to these features only. The fit of the reflectance is similar as in the SFM but is performed in the wavelet space as opposed to on the spectra directly.

Altogether, WAFER yields similar results as other methods for in-situ data, especially at the $O_2A$ band, and performs slightly better for synthetic data, given noise levels are fairly low. Further development is suggested to reduce noise sensitivity, for example by refining the noise thresholding as well as the reflectance fitting function for each retrieval window.

Concluding, we would like to point out the following findings that in their entirety set WAFER apart from other retrieval methods:

(i) WAFER can be applied to arbitrary retrieval windows, as long as the reflectance or multiplicative change to the incoming radiance follows the specified fitting function (in our case a second order polynomial). Additionally, spectral data is explored on a deeper level, because the frequency of the signal (i.e. width of absorption features) is also taken into account. This does not increase the information content but allows for a more thorough analysis with different tools.

(ii) WAFER only puts constraints on the reflectance, which we assume to be a second order polynomial in the employed wavelength ranges. The method is based on the assumption that reflectance and transmittance act multiplicatively on the downwelling radiance. However, this assumption is inherent in all other retrieval methods as well.

(iii) WAFER allows for arbitrary fluorescence spectral shapes. Therefore, changes and shifts in the peaks due to environmental or phenological conditions can be captured. Other retrieval methods induce a coupling or even a degeneracy between fluorescence and reflectance by making simplifying assumptions for the fluorescence spectral shape, approximating the reflectance with the apparent reflectance or reducing the retrieval model (Eq. (1)) to a sum. There is also no need for extensive training datasets specific to each measurement setup.

We recommend further research to exploit capacity and reliability of WAFER for SIF retrievals from airborne and possibly also satellite acquired spectral data. Particularly the capacity of WAFER to retrieve SIF in spectral windows unaffected by atmospheric absorption could yield new possibilities to account for atmospheric effects.

## CRediT authorship contribution statement

**Veronika Oehl:** Conceptualization, Methodology, Software, Formal analysis, Writing – original draft, Visualization. **Alexander Damm:** Conceptualization, Validation, Funding acquisition, Supervision, Writing – review & editing.

## Declaration of competing interest

The authors declare that they have no known competing financial interests or personal relationships that could have appeared to influence the work reported in this paper.

## Data availability

Data will be made available on request.

## Acknowledgments

We acknowledge funding by the SNSF project Fluo4Eco, Switzerland (grant number 197243). We would like to thank Eugénie Paul-Limoges, Michael Niederberger and Bastian Buman for providing the used FloX data and maintaining the instrument and Jennifer Adams for proofreading the manuscript.

## Appendix A. Radiative transfer retrieval assumptions

As the sun is close to being a black body, it emits electromagnetic radiation spectrally following a Planck distribution. This spectrum is equipped with a spectral fingerprint caused by absorptions by atoms making up the solar corona (Fraunhofer, 1817; Kirchhoff, 1860). These absorptions are complemented by molecular absorptions and scattering throughout Earth's atmosphere before the incoming radiation is reflected by the surface, followed by further scattering and absorptions until the radiation is measured with a spectrometer. The upwelling radiation measured is finally composed of the sunlight reflected by the surface, thermal emission and the SIF signal if living vegetation is present.

All of this makes up a complex radiative transfer system where the pathways of the ubiquitous electromagnetic radiation cannot be tracked on the photon level in an analytical way. Deriving a description of these processes using radiation intensities from first principles leads to the integro-differential radiative transfer equation that can only be solved analytically in idealized and simple cases, the Lambert–Beer law being one of them (see Rybicki and Lightman, 1979 for a general introduction). In most applications, this equation is therefore solved numerically.

Although not a formal solution of the radiative transfer equation, radiative transfer is often approximated descriptively as a linear combination of different wavelength dependent reflectance and transmittance factors for the direct and diffuse parts of the radiation. One way of expressing the monochromatic upwelling radiance this way is described for example in Cogliati et al., 2015. A common property of these kinds of expressions is that they can be reformulated in such a way that only two terms remain: one with a linear dependence on the downwelling radiance or solar radiance and the second summarizing all additive fluorescence contributions represented by the variable $F$ in Eq. (1). The factor in front of the downwelling radiance is summarized by the variable $R$ in Eq. (1) and can contain not only the reflectance but also all transmittances. The basic idea of our approach is to retrieve these multiplicative alterations $R$ to a given reference spectrum $s_0$.

## Appendix B. Reference spectra

As our approach simply aims to determine the multiplicative alterations to some kind of reference spectrum, several kinds of references $s_0$ are possible. For this proof of concept we specifically evaluated the case where the downwelling radiance at the top of the canopy is known but in fact, as this approach is developed further, the top of atmosphere solar spectrum or the upwelling radiance from a non-fluorescing target are also viable options.

For retrieval windows with only Fraunhofer lines, a solar reference as presented in Kurucz (2006) alone could be sufficient and would make the retrievals independent of changing atmospheric conditions given that the spectral characteristics (i.e. spectral response function and sampling rate) of the instrument are known and can be applied to the reference spectrum before the retrieval. This is also, where the approach would become feasible for airborne and satellite acquired data as demonstrated in Joiner et al. (2011) for the FLD method and in Joiner et al. (2013) using the SVD method.

**Appendix C. Visual evaluation of the parametrization of the bi-directional reflectance distribution function**

To create our synthetic dataset, we used bi-directional reflectance distribution functions (BRDF) instead of simple scalar reflectances for the different direct and diffuse contributions to the overall radiance. For the BRDFs calculated with SCOPE to be usable within libRadtran, they need to be parametrized in a particular way. This section outlines that the parametrization needed for libRadtran is sufficient to represent the BRDFs calculated with SCOPE. In Fig. C.9 we show BRDFs for different wavelengths and a given configuration of chlorophyll content and leaf area index comparing the SCOPE result to the parametrized BRDF, also showing the relative difference, which is at the percent level for a viewing zenith angle of 0°. This is confirmed by the mean relative deviation, where the mean is taken over all viewing angles, shown in Fig. C.10 over all wavelengths.

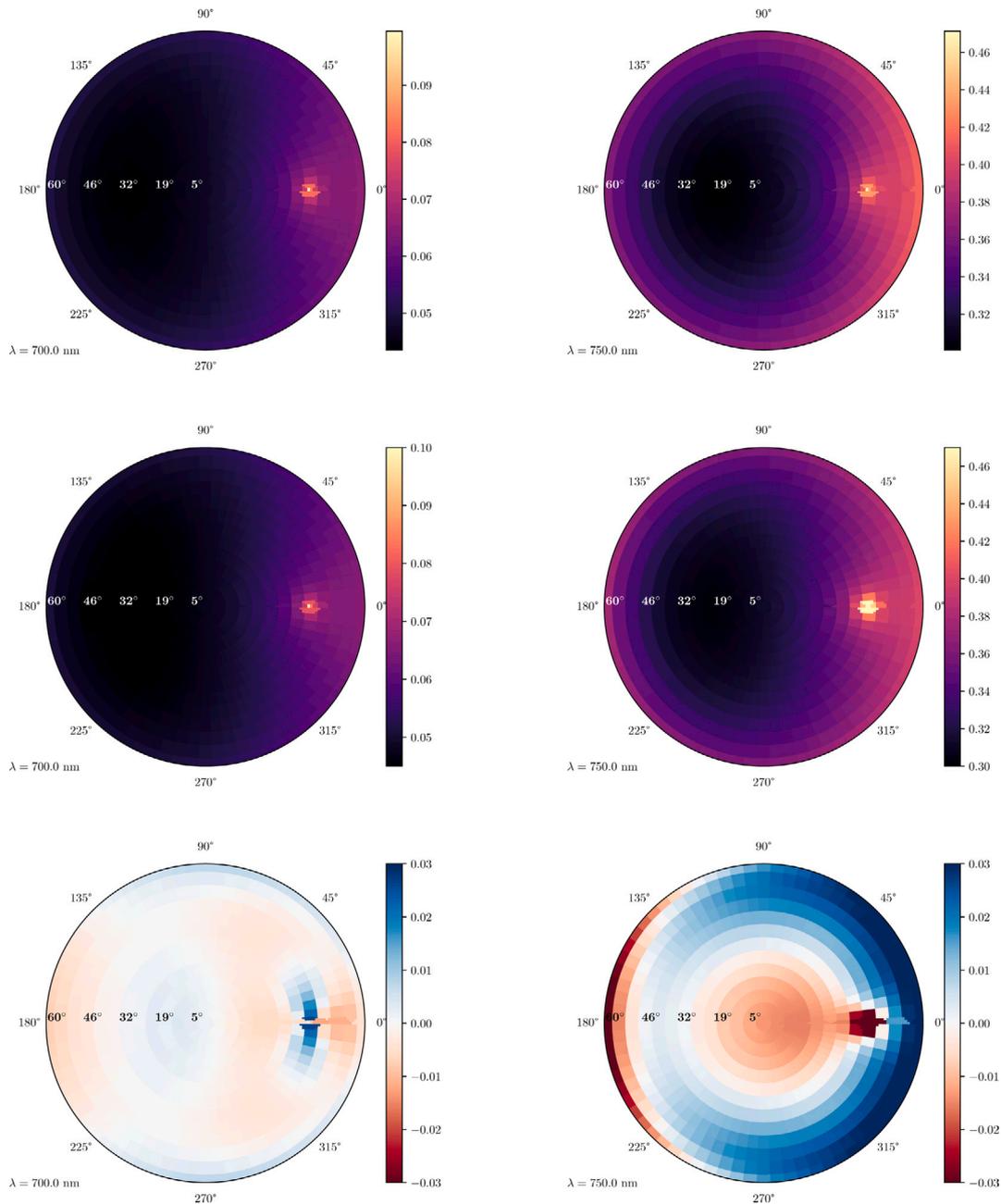

**Fig. C.9.** Bi-directional reflectance distribution functions for different wavelengths (700 nm and 750 nm, left and right column respectively) generated with SCOPE for a chlorophyll content of 50 μg cm$^{-2}$ and a leaf area index of 3 m$^2$ m$^{-2}$ (first row) and their parametrized counterparts using the Ross-Li-Hotspot parametrization (second row). In the third row, the relative difference is shown.

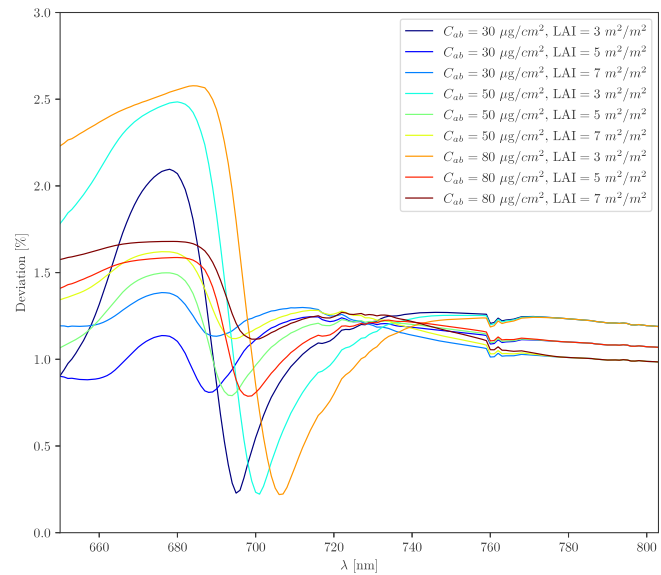

**Fig. C.10.** Mean relative deviation of the parametrized bi-directional reflectance distribution functions (BRDF) from the sampled ones using SCOPE.

### Appendix D. SIF retrieval results from observational data on an overcast day

In this section, we show exemplary fluorescence retrievals on an overcast day as opposed to the sunny days shown in Section 3. Days have been categorized to sunny or overcast by means of the shape of the diurnal cycle of the downwelling radiance. As in Section 3.2, Fig. D.11 shows spectral reflectance and fluorescence retrievals and the mean diurnal difference to the SFM retrievals. In Fig. D.12, fluorescence retrievals with different retrieval methods are compared at characteristic wavelengths.

### Appendix E. Setup of the flox-measurement

See Fig. E.13.

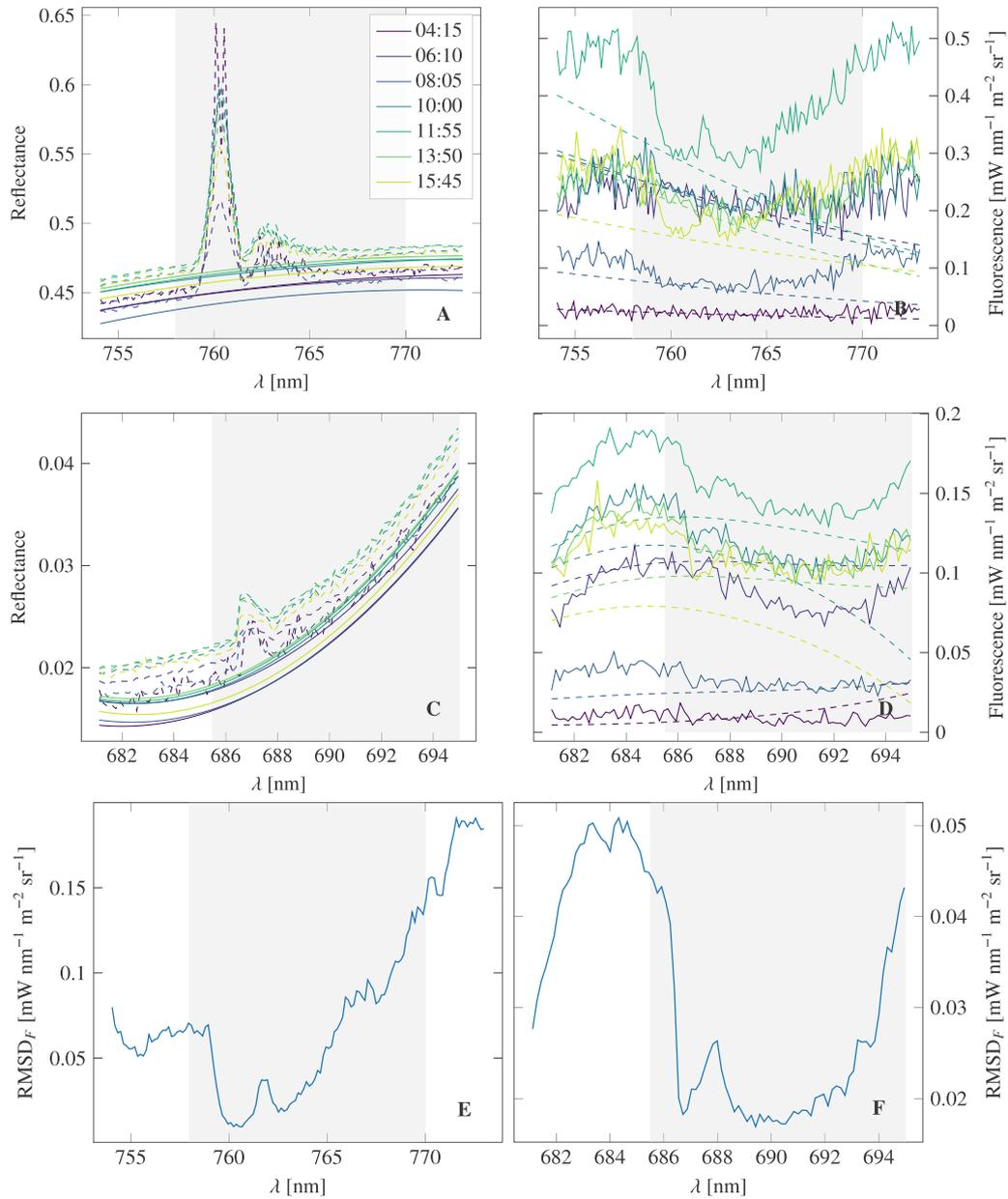

**Fig. D.11.** Reflectances (A and C, solid lines) and fluorescences (B and D, solid lines) retrieved from measured radiance signals comparing reflectance retrievals using WAFER to the apparent reflectance and fluorescence retrievals to fluorescence retrieved with the established spectral fitting method (SFM) (dashed lines). Measurements were taken on 2021-05-11 and results are shown for the retrieval windows including the O₂A band (A and B) and including the O₂B band (C and D). The root mean square differences (RMSD) between the SFM and WAFER for the fluorescence in these windows are shown in panels E and F, where the mean is taken over the entire day. In each panel, the oxygen absorption band is shaded in gray. (For interpretation of the references to color in this figure legend, the reader is referred to the web version of this article.)

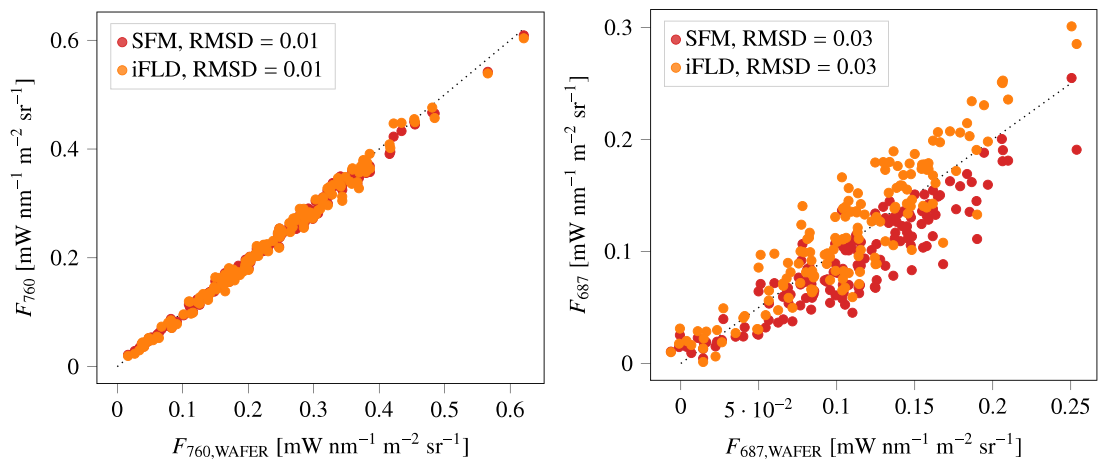

**Fig. D.12.** Comparison of fluorescence retrieved using the improved Fraunhofer line discrimination (iFLD) method and spectral fitting method (SFM) against WAFER at characteristic wavelengths (left panel 760 nm and right panel 687 nm) for all measurements on 2021-05-11. Orange markers represent the iFLD method, red markers the SFM. The dashed lines denote the 1:1 line respectively. The root mean square differences (RMSD) are in $1\,\mathrm{mW\,nm^{-1}\,m^{-2}\,sr^{-1}}$. (For interpretation of the references to color in this figure legend, the reader is referred to the web version of this article.)

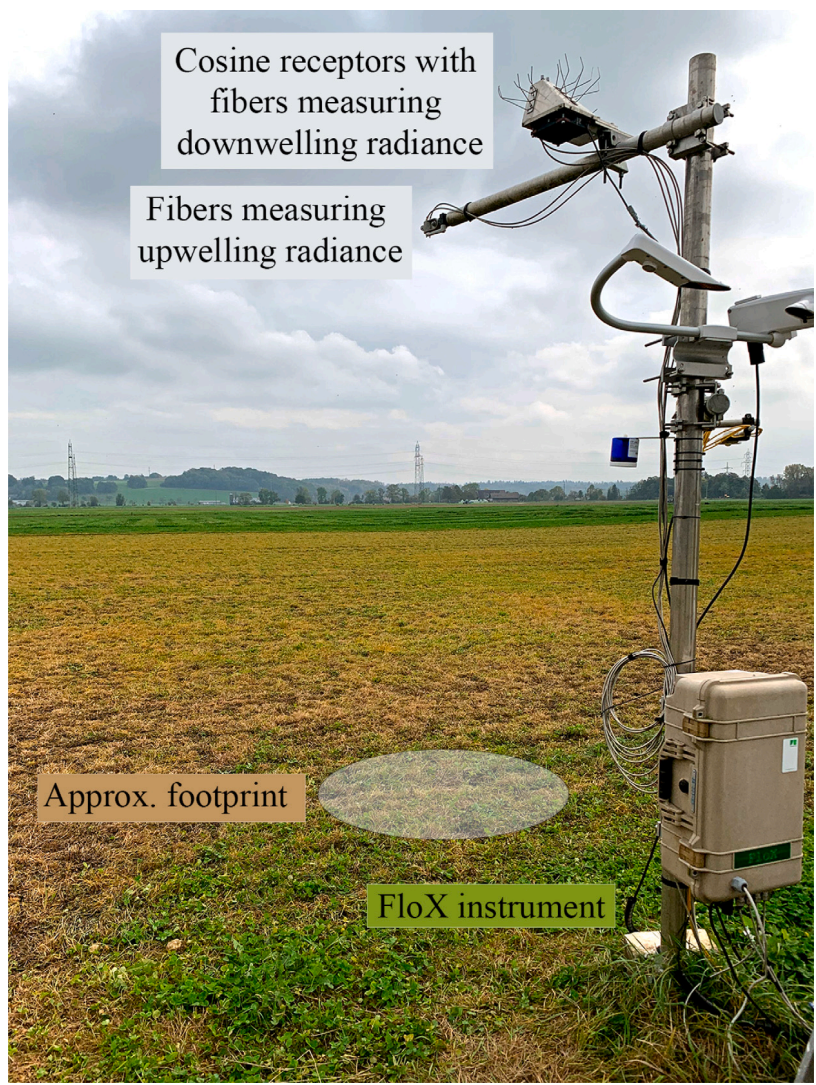

**Fig. E.13.** Setup of the FloX-system at an agricultural site in Oensingen, SO, Switzerland (47.2864° N, 7.7338° E). Picture: Michael Niederberger, annotated.